\documentclass[conference]{IEEEtran}
\IEEEoverridecommandlockouts
\usepackage{cite}
\usepackage{amsmath,amssymb,amsfonts}
\usepackage{algorithmic}
\usepackage{graphicx}
\usepackage{textcomp}
\usepackage[table]{xcolor}
\usepackage{cleveref}
\usepackage{booktabs}
\usepackage{caption} 
\usepackage{subcaption}
\def\BibTeX{{\rm B\kern-.05em{\sc i\kern-.025em b}\kern-.08em
    T\kern-.1667em\lower.7ex\hbox{E}\kern-.125emX}}
\begin{document}

\title{Real-Time Digital Video Streaming at Low-VHF for  Compact Autonomous Agents in Complex Scenes\\
}

\author{\IEEEauthorblockN{Jihun Choi\textsuperscript{\textdagger *1}, Chirag Rao\textsuperscript{\textdagger 2}, Fikadu T. Dagefu\textsuperscript{\textdagger 3}}

\IEEEauthorblockA{\textsuperscript{\textdagger}U.S. Army Research Laboratory, Adelphi, MD 20783\\
\textsuperscript{*}Booz Allen Hamilton Inc., McLean, VA 22102 
}
\textsuperscript\ Email: choi\_jihun@bah.com\\
}

\maketitle

\begin{abstract}
This paper presents an experimental investigation of real-time digital video streaming in physically complex Non-Line-of-Sight (NLoS) channels using a low-power, low-VHF system integrated on a compact robotic platform. Reliable video streaming in NLoS channels over infrastructure-poor ad-hoc radio networks is challenging due to multipath and shadow fading. In this effort, we focus on exploiting the near-ground low-VHF channel which has been shown to have improved penetration, reduced fading, and lower power requirements (which is critical for autonomous agents with limited power) compared to higher frequencies. Specifically, we develop a compact, low-power, low-VHF radio test-bed enabled by recent advances in efficient miniature antennas and off-the-shelf software-defined radios. Our main goal is to carry out an empirical study in realistic environments of how the improved propagation conditions at low-VHF affect the reliability of video-streaming with constraints stemming from the limited available bandwidth with electrically small low-VHF antennas. We show quantitative performance analysis of video streaming from a robotic platform navigating inside a large occupied building received by a node located outdoors: bit error rate (BER) and channel-induced Peak Signal-to-Noise Ratio (PSNR) degradation. The results show channel-effect-free-like video streaming with the low-VHF system in complex NLoS channels.
\end{abstract}

\
\section{Introduction}
 Robust communication in complex scenarios is critical for various applications, including situational awareness and Search and Rescue (SaR) missions for people in emergency situations. Real-time video streaming technology can expedite the rescue operation by sending video data out of the area of danger immediately to a receiving station or rescue teams [1]-[2]. Recently, there has been growing interest in the use of unmanned aerial and ground vehicles for such SaR missions. Since these mobile agents can reach areas that are highly inaccessible or dangerous for human agents, they can substantially improve operational effectiveness.  However, such emergency situations often happen in infrastructure-poor cluttered environments, such as collapsed buildings or tunnels. The wireless communication link is affected by multipath and shadow fading from a multitude of obstacles [3]-[4] which make real-time ad-hoc video streaming in such scenarios unreliable. In order to tackle this challenge, various approaches have been pursued, including signal processing techniques focusing on sophisticated multipath mitigation and channel coding as well as the use of high power [5], which is often undesirable, especially for small mobile platforms. Another approach that has been pursued to address these challenges is utilizing a number of unmanned mobile agents forming a relaying system to disseminate and collect seamless video images in real-time [6]. This comes at an increased cost, and failure of a few nodes could disrupt the communication link.
 
 Recent research on channel characterization of the low Very High Frequency (VHF) band [7]-[8] reveals favorable propagation properties when compared to conventional microwave frequency bands (e.g., Wi-Fi band) especially in Non-Line-of-Sight (NLoS) indoor and outdoor scenarios. Communications experiments have also been carried out with ZigBee radios at 40 MHz using a full-duplex frequency conversion circuit [9]. The results demonstrate that low-VHF operation can provide more reliable and persistent communications in complex scenes, when compared to higher frequency operation, albeit at a reduced bandwidth. 

In this paper, we perform experimental study on a new way of real-time ad-hoc digital video streaming for applications such as unmanned SaR operations to address challenges faced by   existing technology, including sparse and intermittent network connectivity in complex NLoS channels. Our approach introduces a compact, mobile, low-power, low-VHF radio system integrated on an autonomous agent enabled by recent advances in small efficient antennas [10]-[11], which can potentially provide much larger and reliable coverage with a much smaller number of unmanned mobile agents. Of course, there is a trade-off among system size, data rate, and video quality, due to the use of low frequency systems. Thus, appropriate parameters such as data-rate, bandwidth, video bitrate, and frame rate, adequate for applications such as SaR missions, need to be determined. 

We employ a pair of commercial off-the-shelf (COTS) Software Defined Radios (SDR) to transmit and receive a real-time video stream from a webcam. We carefully design baseline system blocks, including transcoding process for the source video stream and the digital modulation/demodulation process for radio communications. In order to evaluate performance of the low-VHF video streaming mobile system in complex NLoS channels, we quantitatively characterize digital communication and video quality with data collected in such environments. The performance metrics used are Bit Error Rate (BER) and channel-induced Peak Signal-to-Noise Ratio (PSNR) degradation. The experiment results support the proposed concept of reliable and robust live video streaming in complex scenes, enabled by the unmanned compact low-VHF radio systems.

The rest of the paper is organized as follows. Section II describes the overall system, including a small unmanned ground vehicle (UGV), software and hardware integration of SDRs and a highly miniaturized efficient low-VHF antenna as well as the design of wireless digital communication framework for low-VHF video streaming. Section III presents the experimental scenario and setup, followed by performance evaluation which includes measurement results representing wireless communication quality and corresponding video quality.  

\begin{figure}[t]
\centerline{\includegraphics*[width=3.40in]{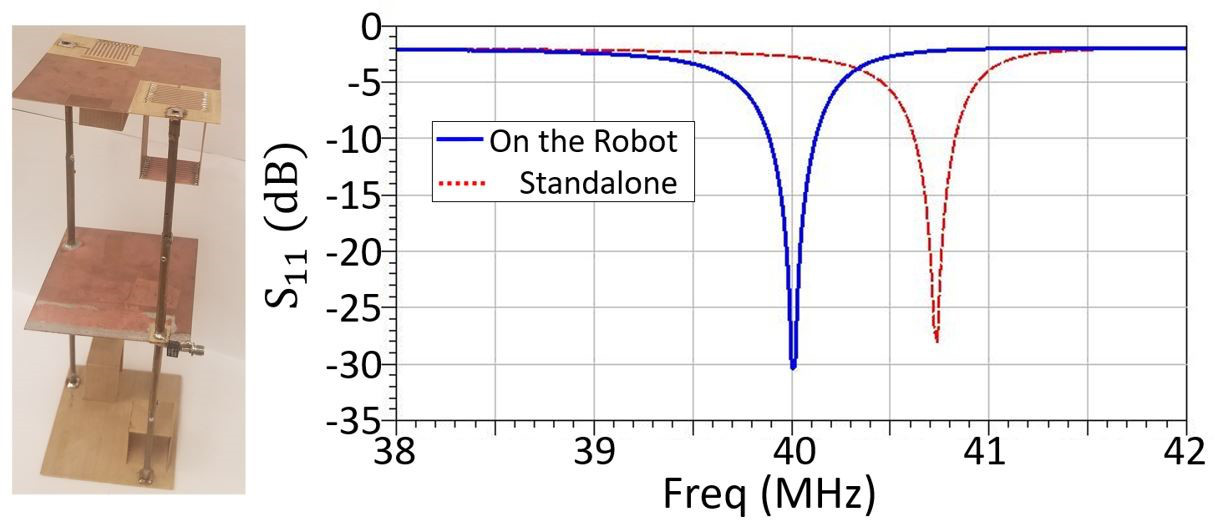}}
\caption{{\bf Left:} The fabricated highly miniaturized low-VHF antenna and {\bf Right:} its measured input impedance are shown.}
\label{fig:fig_1}
\end{figure}
\
\section{Low-VHF System For Small Robotic Platforms}
A major challenge of the proposed work is establishing a wireless mobile system which has a small form factor operating at low power within the low-VHF band. Because antenna size is proportional to wavelength, the physical dimensions of conventional low-VHF antennas are prohibitively large. Therefore, an efficient miniature antenna design is critical, as it enables low-power operation as well as compactness of the overall system. To realize a flexible test-bed of real-time digital video streaming for SaR-type missions, we also utilize SDRs, which facilitate the implementation and experimentation of the wireless digital communication system, along with a mobile robot enabling autonomous exploration.
\

\subsection{ An Efficient Miniature Low-VHF Antenna}

Electrically small efficient antennas are key enabling components in compact and low-power wireless communication systems at the low-VHF band. In general, antennas at this band are designed to have heights several meters long (e.g., a quarter-wavelength monopole at 40 MHz). Furthermore, antenna bandwidth predominantly determines the system bandwidth for low-VHF wireless communications. In [10]-[11], different types of miniature antennas are introduced for low-VHF applications. Narrow bandwidths of the antennas, however, limit their application to low data rate communications. For the proposed application, the antennas may not be compatible allowing for minimum bandwidth required to transmit real-time video [12]. Antenna gain besides bandwidth is also an important aspect for enabling low-power operation. It should be mentioned that small antenna design simultaneously satisfying broad bandwidth and high gain is infeasible due to the trade-off between antenna size and performance. 

The miniature antenna in [10] is designed to have much higher radiation efficiency compared to the similar-sized antennas, but its bandwidth is relatively narrow due to the trade-off. For the proposed application, a modified version of the antenna [11] is utilized. Fig. \ref{fig:fig_1} shows a picture of the fabricated antenna and its matching performance. This folded dipole-based antenna consists of optimized rectangular air-core inductors with high Q factors and capacitive loadings connected through metallic poles that are the main radiating elements. The overall dimension of this antenna is 10 cm $\mathrm{\times}$ 10 cm $\mathrm{\times}$ 30 cm which corresponds to ${\lambda }_0$/75 $\mathrm{\times}$ ${\lambda }_0$/75 $\mathrm{\times}$ ${\lambda }_0$/25 in terms of its electrical length ${\lambda }_0$ at 40 MHz. A fractional bandwidth of the antenna for a voltage standing wave ratio of 2:1 is 0.57 \%. The antenna provides an omnidirectional radiation pattern with a gain of -10 dBi reasonably suitable for low-power communication experiments with a mobile robotic platform. 

\begin{figure}[t]
\begin{center}
\begin{subfigure}{\linewidth}
\includegraphics*[width=\linewidth]{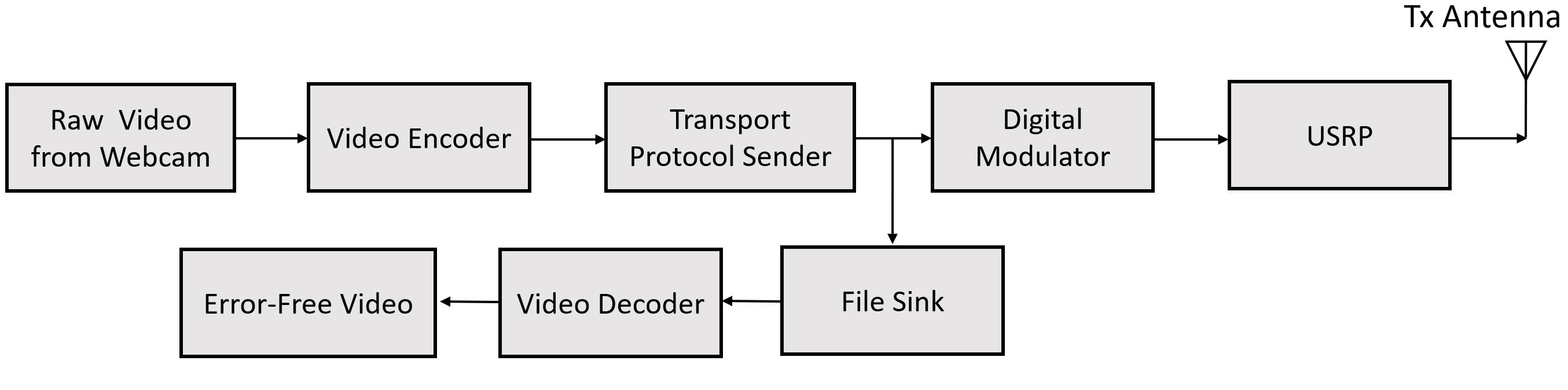}
\caption{}
\end{subfigure}
\\
\begin{subfigure}{\linewidth}
\includegraphics*[width=\linewidth]{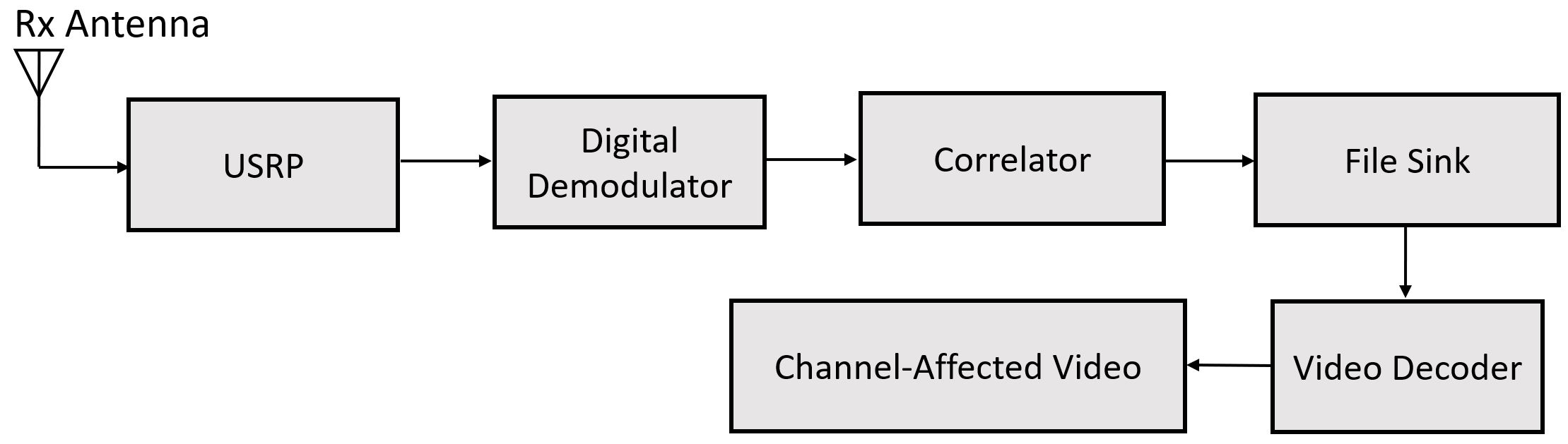}
\caption{}
\end{subfigure}
\end{center}
\caption{A flow graph of a framework developed for the proposed low-VHF digital video streaming: (a) real-time video transmitting and (b) receiving processes.}
\label{fig:fig_2}
\end{figure}

\subsection{ Software Defined Radios}

Software-defined radios are widely used for wireless applications such as data, voice, and video communications due to their effectiveness in flexible spectrum management and communications systems prototyping. For our radio experiments, we use a COTS SDR, a Universal Software Radio Peripheral (USRP) X310 [13], with a full-duplex transceiver daughterboard, a UBX 160, which performs baseband processing of signals, digital-to-analog conversion (DAC), analog-to-digital conversion (ADC), frequency up/down conversion, and signal amplification and filtering. The maximum transmit power of an X310 transmitter is 20 dBm and the noise figure of a receiver is less than 3 dB at the low-VHF band. 

To create and control the software-defined radio system, we use the GNU Radio software suite with the USRP hardware driver (UHD) [14]. Fig. \ref{fig:fig_2} depicts a flow graph of a framework developed for the proposed low-VHF digital video streaming. The transmitter receives video frames via User Datagram Protocol (UDP) packets and appends a custom header to each packet. The custom header consists of a 64-bit preamble sequence used for packet detection at the receiver, a 2-byte indicator of the number of modulated bits per symbol, a 2-byte payload length indicator, and a 2-byte packet sequence number.  The combined header and payload are Binary Phase Shift Keying (BPSK) modulated, passed through a Root Raised Cosine (RRC) filter, and transmitted through the USRP. On the receive side the modulated signal is received through another USRP in the form of in-phase and quadrature (I/Q) samples. The samples are processed by the BPSK demodulation block, which first attempts to track the carrier, then applies fine-frequency correction and RRC filtering, and finally demodulates the signal. The demodulated bits are passed through a correlator block, which detects packets by correlating against the 64-bit packet preamble. The detected headers of the packets are then parsed and the payload and header information are stored for playback and analysis.  

\subsection{ A Video Streaming System on a Small Mobile Robot}

\begin{figure}[t]
\centerline{\includegraphics*[width=\linewidth]{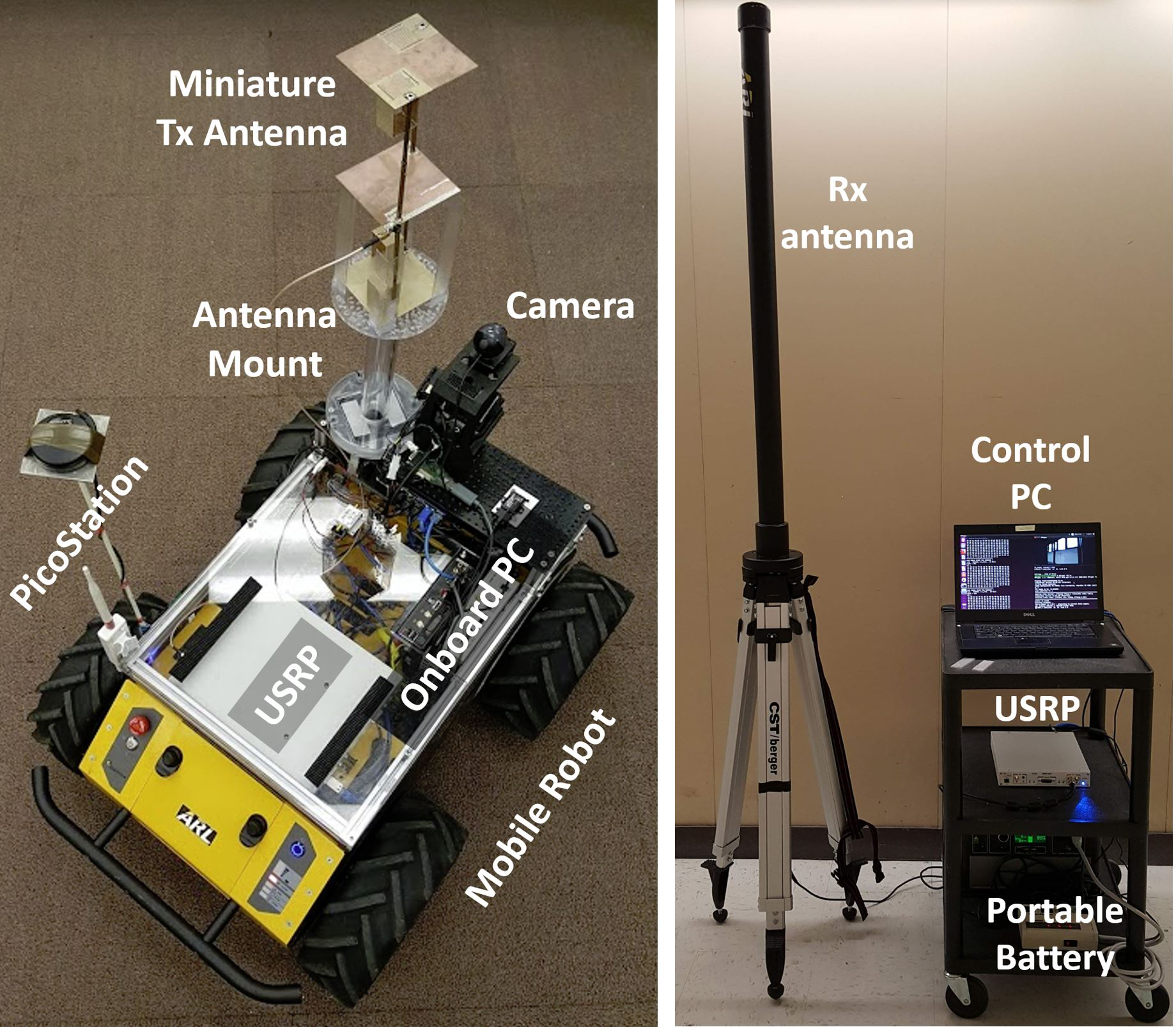}}
\caption{{\bf Left:} A compact low-VHF video streaming system integrated on a small autonomous mobile agent and {\bf Right:} a receiving station to play back real-time video.}
\label{fig:fig_3}
\end{figure}

The aforementioned antenna and SDR are integrated on a small UGV for real-time low-VHF digital video streaming. Fig. \ref{fig:fig_3} {\bf (Left)} illustrates the fully integrated system on the robot. The miniature antenna is further tuned to remove the resonant frequency shift caused by the polycarbonate antenna mount (see Fig. \ref{fig:fig_1} {\bf (Right)}). The antenna performance with the  system setup is characterized to confirm that there are no coupling effects between the antenna and nearby objects, including the robot. The various hardware components, including a webcam, the USRP, Picostation (for short range remote control of the robot), and the robot itself are controlled by an on-board PC through the robot operating system (ROS). A battery pack of BB-2590 and appropriate voltage regulators are mounted on the UGV to provide required power for the components. On the receive side (see Fig. 3 {\bf (Right)}), we use a SDR, a control PC, a high capacity portable battery, and a short dipole antenna with length 1.25 meters (= ${\lambda }_0/6$ at 40 MHz in terms of electrical length). 

For low-VHF video streaming, we use a low-resolution webcam whose maximum frame rate is 15 frames per second (FPS) at 640 $\mathrm{\times}$ 480 pixels [15]. A video source from the webcam is encoded through GStreamer [16], to H.264 which is a commonly used video compression standard [17]. The digitally compressed video is encapsulated into the aforementioned UDP packets and sent to GNU Radio. The H.264 encoding parameters such as resolution, frame rate, and data rate need to be properly determined, as they critically affect the video quality. Since system bandwidth for the proposed application is limited by the antenna bandwidth, high video compression is required. Based on preliminary experimental investigation, acceptable video quality for SaR-type applications [18] compatible with the available bandwidth is determined. The parameters used for the proposed application are tabulated in Table \ref{tab1:tab_1}. 

\
\section{ Experiments in Complex Scenes }

\begin{table}[t]
\caption{SYSTEM PARAMETERS}
\begin{center}
\begin{tabular}{|c|c|c|c|}
\hline
{\bf Resolution} & {\bf Frame Rate} & {\bf Video Bit Rate} & {\bf Sampling Rate} \\\hline\hline
320$\times$240 pxs & 15 FPS & 300 / 500 kbps & 500 kbps \\\hline
\end{tabular}
\label{tab1:tab_1}
\end{center}
\end{table}

\begin{figure}[t]
\centerline{\includegraphics*[width=\linewidth]{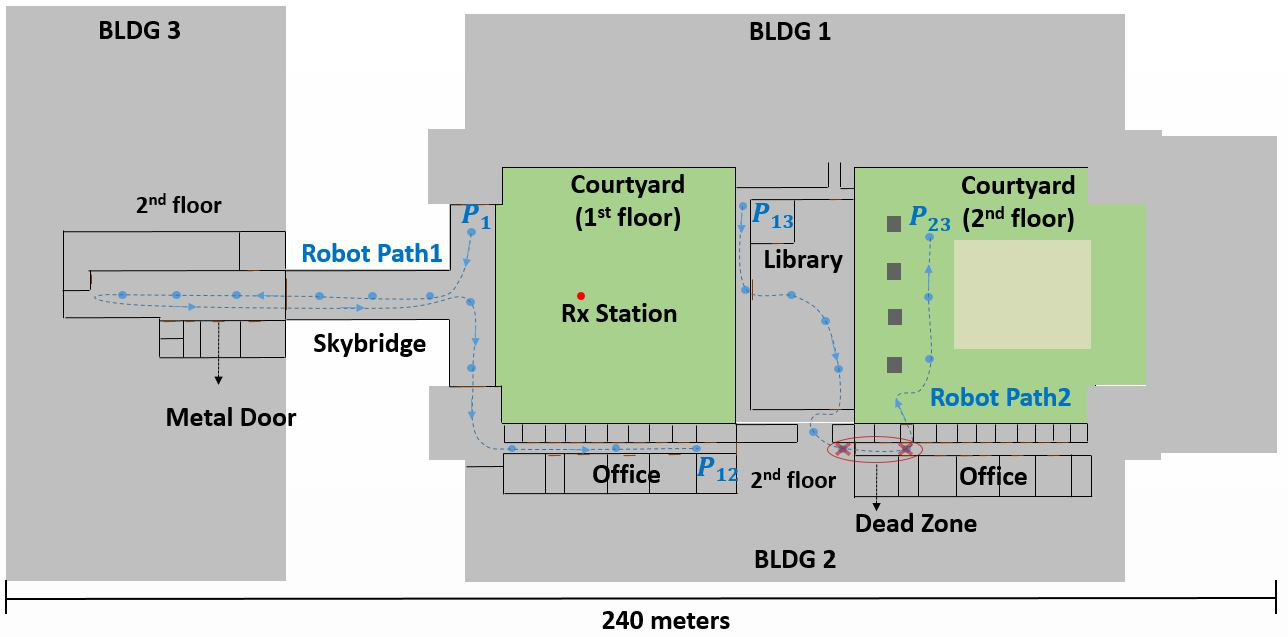}}
\caption{Complex measurement scenarios: a combination of NLoS multi-floor indoor-to-outdoor and NLoS outdoor scenes. A receiving station is positioned on the 1st floor courtyard and a real-time video streaming mobile robot traverses along the paths indicated.}
\label{fig:fig_4}
\end{figure}

 We perform experiments with the system in various realistic complex channels. Fig. \ref{fig:fig_4} illustrates one of the measurement scenarios in NLoS multi-floor indoor-to-outdoor and NLoS outdoor-to-outdoor channels. The UGV with the video streaming system is maneuvered along the various paths shown inside the building. The receiving station is positioned in the center of the first floor courtyard surrounded by an occupied 5-story building. 
 
 \begin{figure}[t]
\begin{center}
\begin{subfigure}{\linewidth}
\includegraphics*[width=\linewidth]{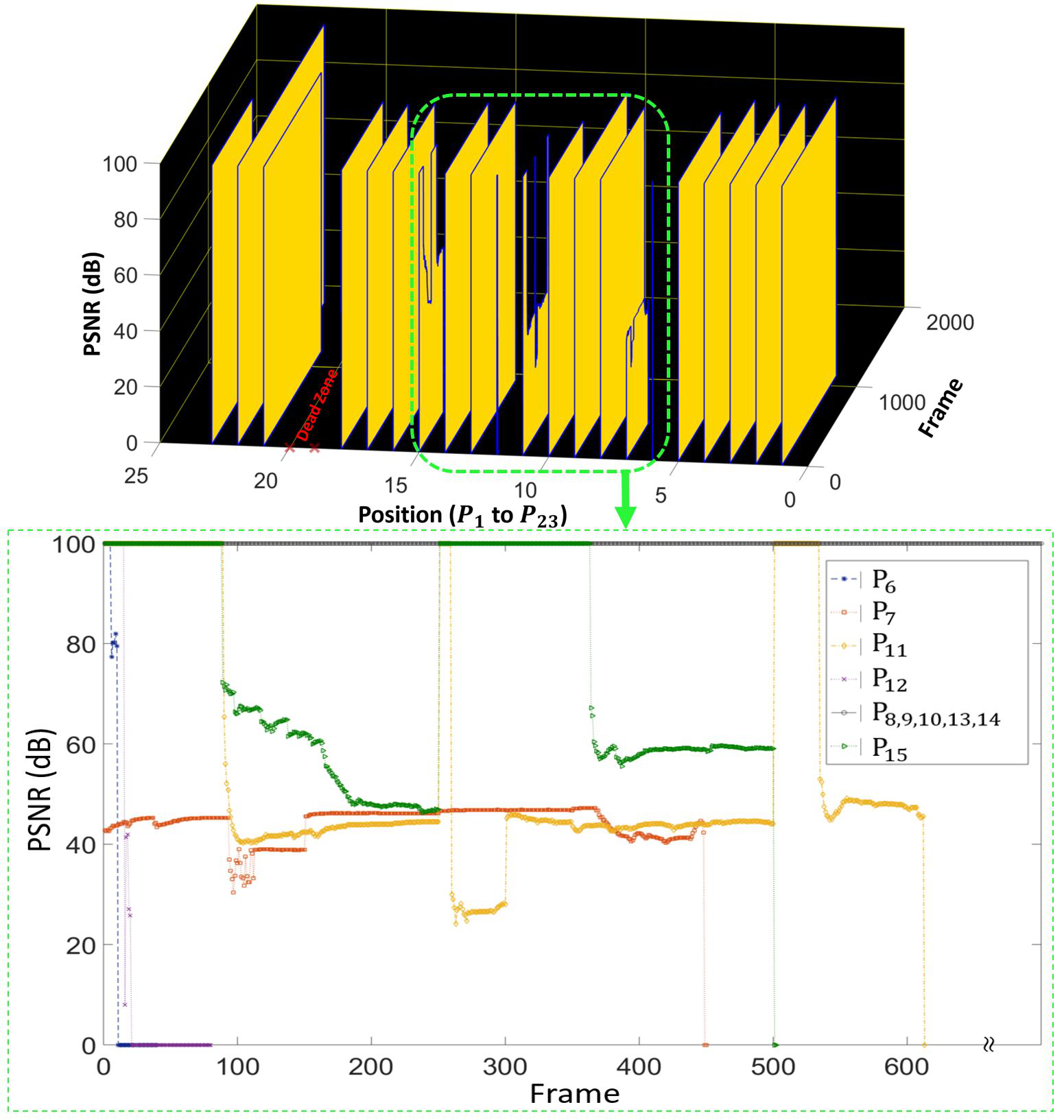}
\caption{}
\end{subfigure}
\\
\begin{subfigure}{\linewidth}
\includegraphics*[width=\linewidth]{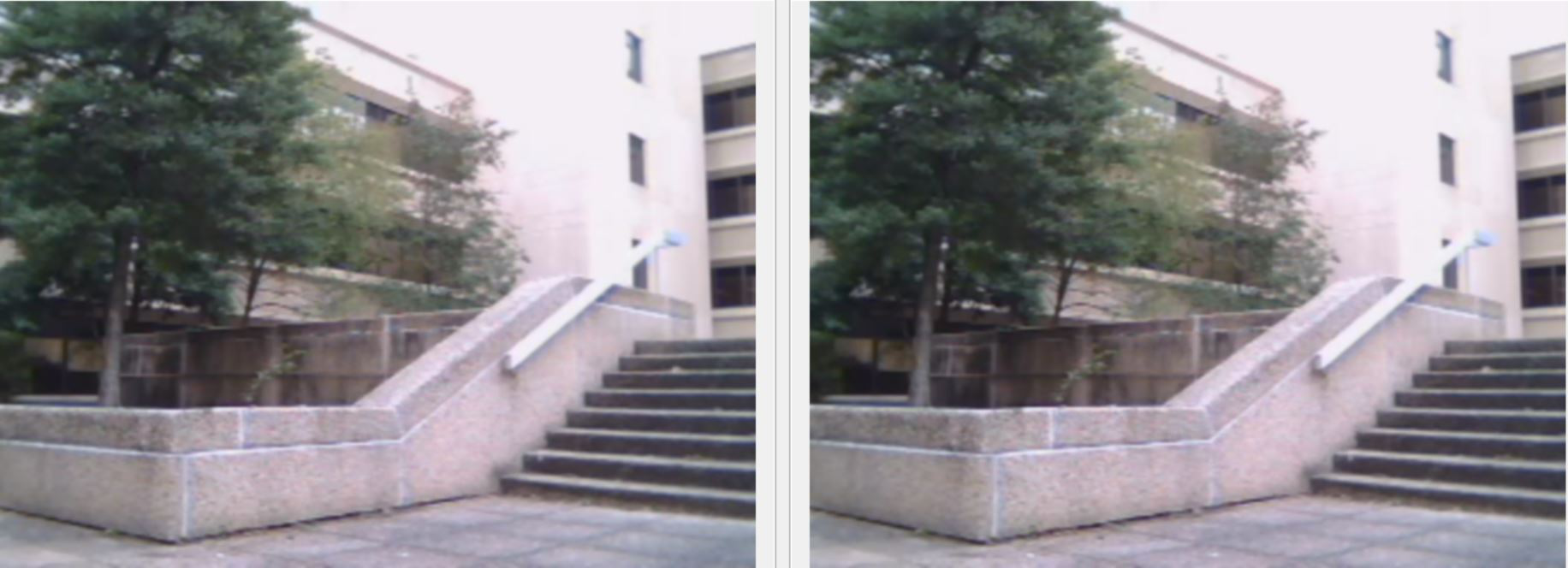}
\caption{}
\end{subfigure}
\\
\begin{subfigure}{\linewidth}
\includegraphics*[width=\linewidth]{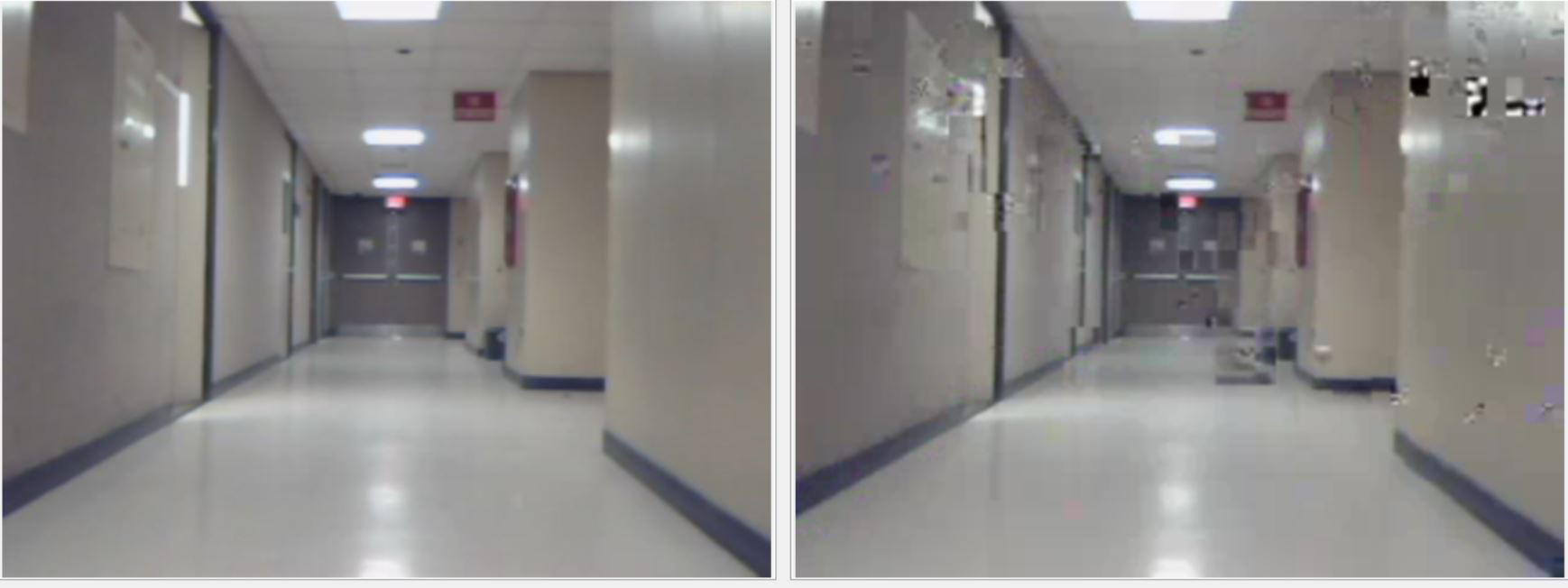}
\caption{}
\end{subfigure}
\end{center}
\caption{(a) PSNR on each frame of the recorded videos at the points marked in Fig. 4 and frame-by-frame comparison for subjective video quality assessment: {\bf Left} (b) \& (c) are error-free frames and {\bf Right} (b) \& (c) are channel-affected frames with PSNR of 100 dB and 27 dB, respectively.}
\label{fig}
\end{figure}

Note that this particular scenario has a high degree of complexity in terms of radio wave propagation. First, the height difference between the transmitter and the receiver is about 6 meters, affecting the angle of antenna radiation and the effect of ground reflections that may degrade wireless communication quality. Second, there is reduced penetration and increased fading even at lower frequencies. To understand the complexity of this specific scenario, we first carry out path-loss comparison measurements at two different frequency bands with a different combination of antennas (low-VHF vs. Wi-Fi). Unlike other complex environments in [7]-[9], the path-loss difference between the two bands is not as high as other complex environments previously tested where we observed well above 30 dB improvement at low VHF. Furthermore, a significantly higher level of received signal fluctuation at Wi-Fi band is observed. This is because the buildings are composed of a large amount of large metallic components (e.g., all walls between offices and doors are made of large metal sheets). We select this challenging scenario since it has representative channel conditions for a SaR-type application. We carry out quantitative performance evaluation of real-time digital video streaming with the compact low-VHF radio systems that operate at low power ($<$ 20 dBm). 

\subsection{ Video Quality Measures}

A direct way to evaluate effectiveness of the proposed video streaming approach is to measure video quality as perceived video degradation, subjectively or objectively. For quantitative assessment, we leverage a commonly used objective video quality model in practice, PSNR, which is calculated between every frame of error-free and the channel-affected video signal for our experiment (see Fig. 2). The PSNR is defined as\\
\begin{equation} \label{GrindEQ__1_} 
PSNR\ (dB)=10\cdot {log}_{10}\frac{{\max_c}^2\cdot w\cdot h}{\sum^{w,h}_{i=1,\ j=1}{{(X_{i,j}-Y_{i,j})}^2}}, 
\end{equation} \\
where $\max_c$ is the maximum possible color components of the image, \textit{w} is the video frame width, \textit{h} is the video frame height, \textit{X} is an error-free \textit{w}$\times $\textit{h} video frame and \textit{Y} is its corresponding frame affected over a wireless channel. For our experiment, $\max_c$ corresponds to 255 for the image size of 320 $\mathrm{\times }$ 240 with a bit depth of 8 bits. 

 We utilize open-source software [19] to characterize the frame-by-frame PSNR comparison of the recorded video files at the marked points indicated in Fig. 4. Fig. 5(a) shows calculated PSNR for each frame of each video sample. The streamed video samples are collected at the marked points while the mobile robot rotates around. Fig. 5(b)-(c) depicts an example of the frames captured for subjective quality assessment in addition to PSNR; Fig. 5 {\bf Left} (b) and (c) are the error-free frames and Fig. 5 {\bf Right} (b) and (c) are the channel-induced frames with PSNR of 100 dB and 27 dB, respectively. Analysis of the results arises some observation worth mentioning: 1. At most points the received videos show no signs of being degraded by channel effects, and 2. in a given received video, there are a few frames that exhibit high PSNR degradation or no received frames (dead zone in Fig. 4) in the selected complex scene. These observations give rise to further investigation of correlation between digital communications and real-time video streaming quality over complex wireless channels, which is covered in the following section.

\subsection{ Digital Communcation Quality Measures}

In order to study the effect of the digital communication quality on perceived real-time video degradation, we measure BER and correlate it with average PSNR (APSNR). To confirm that occurrence of bit errors is not from the system (e.g., video codecs, imperfect hardware units) but solely due to the wireless communication channel, first we perform BER measurements for the low-VHF video streaming systems in which Tx and Rx ports of the radios are connected with a wire. 

\begin{figure}[t]
\centerline{\includegraphics*[width=\linewidth]{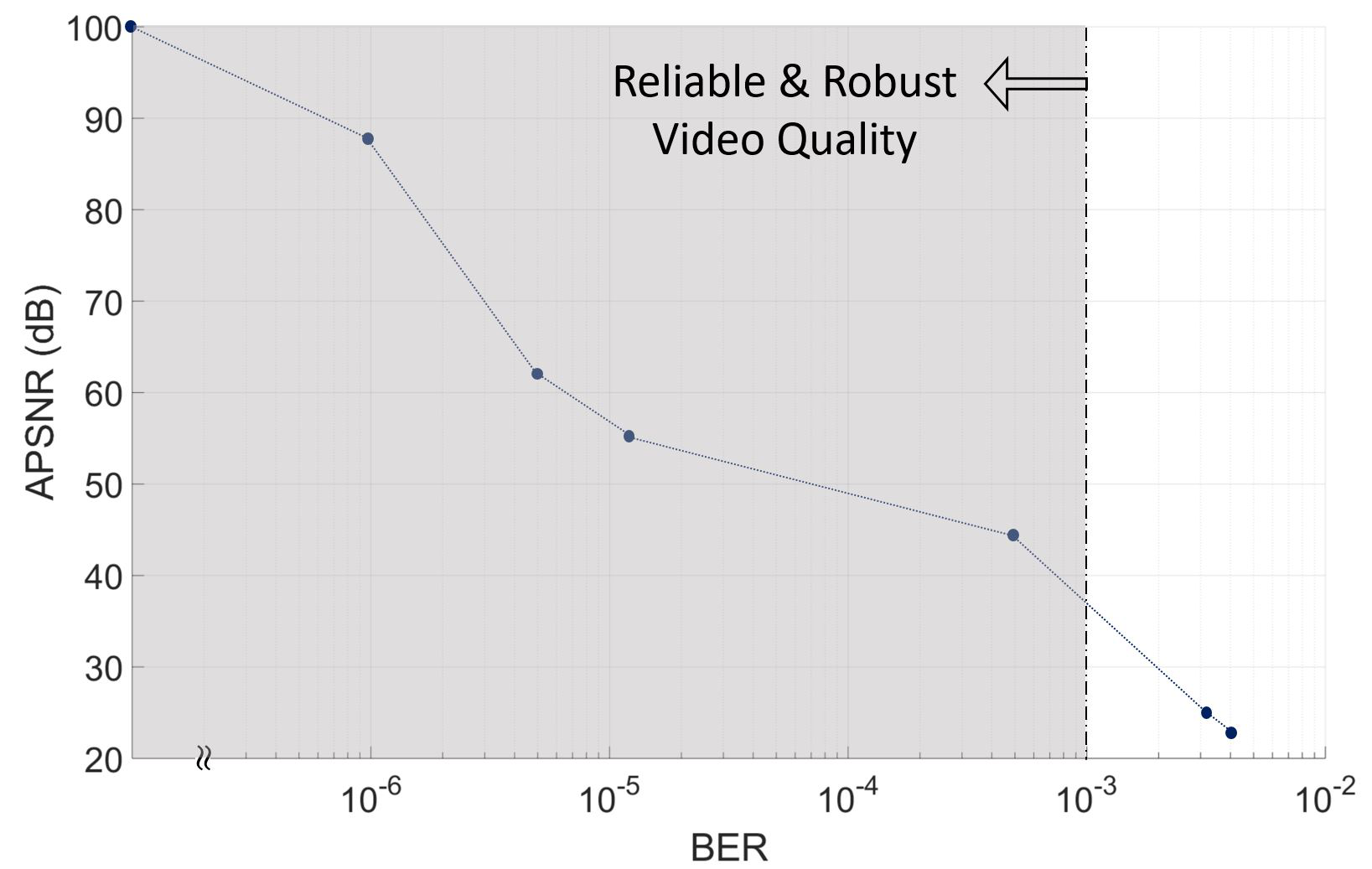}}
\caption{Average PSNR (APSNR) as a function of BER measured at the marked points in Fig. \ref{fig:fig_4}. For the characterization, transmitted and received video data packets are matched by their packet sequence numbers. Here, 100 dB of APSNR corresponds to zero BER.}
\label{fig:fig_6}
\end{figure}

Next, over wireless channels we perform the measurements along the robot paths highlighted in Fig. \ref{fig:fig_4} and the results of APSNR as a function of BER are shown in Fig. 6. To measure APSNR and BER, transmitted and received video data packets are matched by their packet sequence numbers, and the corresponding payloads are compared byte-by-byte for bit errors. At the points where the flawless video playback is achieved, no bit error is observed for video stream with a length of more than 1$\mathrm{\times }{10}^7$ bits per sample. In the dead zone as shown in Fig. 4, most packets are dropped and few detected packets were not enough to retrieve any meaningful video frames with our receiver processing. Furthermore, at the boundary region where a sequence of packets is partially dropped, the measured BER ranges from $8.58\times {10}^{-7}$ up to $\mathrm{3.20}\times {10}^{-3},$ which corresponds the PSNR value ranging from 8 dB up to 100 dB. Based on the measurement results, acceptable PSNR values correlating with perceived video quality effective for robust SaR applications would be above 30 dB where BER is less than $1\times {10}^{-3}$, which is similar to criteria in [20]. 

\noindent It should be mentioned that signal processing has been kept to a minimum in this study. Complex techniques that would be used for channel equalization, dynamic control, forward error correction, and coarse frequency offset compensation between radios have been omitted. This fact puts our system in direct contrast with similar video streaming platforms that operate at higher frequencies for wireless digital communication, where such techniques are necessary for successful communication in such complex environments. This also suggests that further improvement of video quality with high reliability and robustness over longer range and wider coverage can be expected. 

\
\section{Conclusion}

We have experimentally investigated real-time digital video streaming with unmanned, compact, low-power, low-VHF radio systems aimed to strengthen SaR operations. Compactness and low-power operation of the systems are enabled by a recently developed efficient miniature antenna. Digital communications to transmit and receive a video stream over a wireless channel are established using SDRs integrated into a small mobile robot. Performance characterizations of the proposed low-VHF systems with respect to digital communication and video quality were conducted in various NLoS environments. Without complicated signal processing techniques, a channel-effect-free-like quality of digital video streaming was achieved with low-power ($<20\ $dBm) over a multipath-rich channel. Hence, the proposed systems can be very effective for applications where reliability and robustness are critical. We are currently working on bandwidth-enhanced low-VHF video streaming with a recently designed novel antenna [21]. The antenna facilitates further size reduction yet bandwidth enhancement of the overall system, breaking the aforementioned trade-off between size and performance.\\\

\bibliographystyle{ieeetr}

\end{document}